\documentclass{aa}
\usepackage{epsf}

\def\etal{{et~al.\,}}
\def\astl{Astron. Lett.}    
\def\aap{A\&A}          
\def\apj{ApJ}           
\def\sval{Sov. Astron. Lett.} 

\def\4u{4U1724$-$307}

\begin{document}

\thesaurus{13(08.14.1;          
              10.07.3 Terzan 2; 
              13.25.1;          
              13.25.5)}         

\title {An X-ray burst with strong photospheric radius expansion\\
observed from the source 4U1724$-$307 in Terzan\,2}

\author {S.\,V.\,Molkov\inst{1,2}, S.\,A.\,Grebenev\inst{1,2}, and
         A.\,A.\,Lutovinov\inst{1}} 

\institute{
Space Research Institute, Russian Academy of Sciences, 
Profsoyuznaya 84/32, 117810 Moscow, Russia 
\and
Max-Planck-Institut f\"ur Astrophysik, Karl-Schwarzschild-Str. 1,
85740 Garching, Germany}

\offprints{S.\,V.\,Molkov $<\!$msv@hea.iki.rssi.ru$\!>$}

\date{Received November 15, 1999; accepted April 19, 2000}

\titlerunning{An X-ray burst with strong photospheric radius
   expansion observed from Terzan 2} 
\authorrunning{Molkov et al.}
\maketitle 

\begin{abstract} 
We present results of the RXTE observations of an extremely intense
X-ray burst detected from the source \4u\ in the globular cluster
Terzan~2. The burst profile was complex consisting of two precursors and
a long primary peak. During the first (strong) precursor the source
luminosity, $L\sim3.6\times10^{38}\, \mbox{\rm erg s}^{-1}$, was
comparable with that measured in the main event. The structure of the
profile, its dependence on energy and observed spectral evolution
indicated strong photospheric expansion of a neutron star. The effective
temperature and radius of the photosphere were estimated from the black
body model at different stages of expansion, their correlations and
specific features of the profile were analyzed and discussed.

\end{abstract}

\section{Introduction}
Many bright X-ray sources in globular clusters are X-ray bursters. The
source in Terzan~2, the metal-rich cluster located in the Galactic
center field, is likely the most peculiar one. In 1975 September a
strong ($\sim 1.5$ Crab in a peak flux) and long ($\ga 300$\,s) burst
was detected from the region by \mbox{OSO-8} (\cite{swank77}).
\cite{grind78} was the first who associated it with Terzan~2 and noted
that the source {4U1722$-$30} could be its persistent counterpart.
The source position in the {\it Uhuru\/} catalogue was found 
affected by contamination of nearby sources. After correction for this
effect the coincidence in positions of the source, the burster and
Terzan~2 became obvious. Since that time the source is known as \4u.

The identification of the burster with the cluster was confirmed
during the second X-ray burst detected by the {\it Einstein\/} HRI in
1979 March (\cite{grind80}). The source was localized to be within the
cluster core of $\sim6.5$\arcsec\ in radius (\cite{grind84}).
Recent ROSAT HRI images (\cite{mereg95}) also revealed a single 
source less than 2\arcsec\ off the position determined by {\it
Einstein}.

\begin{figure}[b]
\vspace{-3mm}
 \epsfxsize=92mm
 \hspace{-2mm}{\epsffile{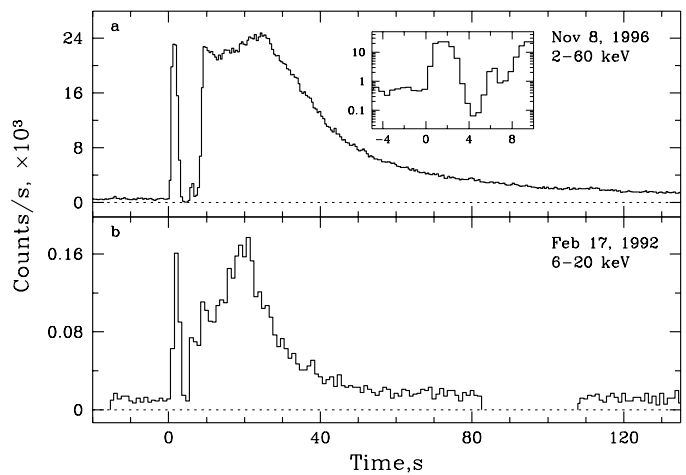}}

\caption{\small Temporal profiles of the bursts detected by the RXTE PCA
({\it a}) and the {\it Granat} ART-P ({\it b}) in the broad energy
bands. Gap in the ART-P curve between 80 and 110 s is due to data
transmission from the temporal buffer to the on-board memory.}
\label {totcur}
\vspace{-0.7mm}
\end{figure}
\noindent
Since 1979 only two new X-ray bursts were reported from the source in
addition to the two already mentioned. The one was detected by the
\mbox{ART-P} telescope on board {\it Granat\/} in 1992 February
(\cite{greb99}) and another -- by the MECS spectrometer on board
\mbox{BeppoSAX} in 1996 August (\cite{guain98}). All the four reported
bursts were extremely bright with short rise and long decay time scales
that is a characteristic of the type~I bursts. For all the bursts the
complex spectral behaviour was observed during the initial phase along
with the gradual spectral softening during the decay phase. In spite of
such similarity some features in the burst profile detected with
\mbox{ART-P} 
(see Fig.\,\ref{totcur}) made that event to be different from the other
ones. In particular, a strong narrow precursor was detected a few
seconds before the beginning of the primary peak that was a direct
evidence for the nearly Eddington luminosity and quick photospheric radius
expansion of a neutron star in \4u. In this letter we present and
discuss the first  results of timing and spectroscopy for another
exceptionally bright X-ray burst with strong photospheric radius
expansion which was detected from \4u\ by the {\it Rossi X-ray Timing Explorer}.
\section{Observations}
In 1996-1998 \4u\ was observed by RXTE many times with a total
exposure exceeding 330~ks but with the only X-ray burst detected. This
event occurred on November 8, 1996 during the first orbit of the
observation started at 6:58 a.m. (UT).
All the five units of the proportional counter array (PCA) 
were operated that time.

The burst profile in the 2-60 keV energy band (which is presented in
Fig.\,\ref{totcur} and discussed below) is based on the Standard-1 mode
data taken from the PCA xenon layers. These are the data without energy
resolution. For the spectral analysis the Burst Catcher mode data were
used recorded in 64 energy channels with 2-ms time resolution.  This
mode, adapted to high count rate, was unfortunately switching off all
the time when the rate fell below $\sim 2\times 10^3$ cts/s, e.g.
during the period taken place between the precursor and primary event.
The data in the Science Event mode could not be used because of strong
memory restrictions and long (8~s) ``readout'' time.  To measure the
spectra just before and in $\sim 300$ s after the burst we used the
Standard-2 mode data with the 16-s integration time.

Our research is focused on the X-ray burst only. Spectral and timing
properties of the source persistent emission were investigated with RXTE
earlier and reported by Olive et al. (1998) and Barret et al. (1999a).

\section{Profile of the burst and precursors}
\begin{figure}[t]
 \vspace{0.6mm}
 \epsfxsize=91.5mm
 \hspace{-2.0mm}{\epsffile{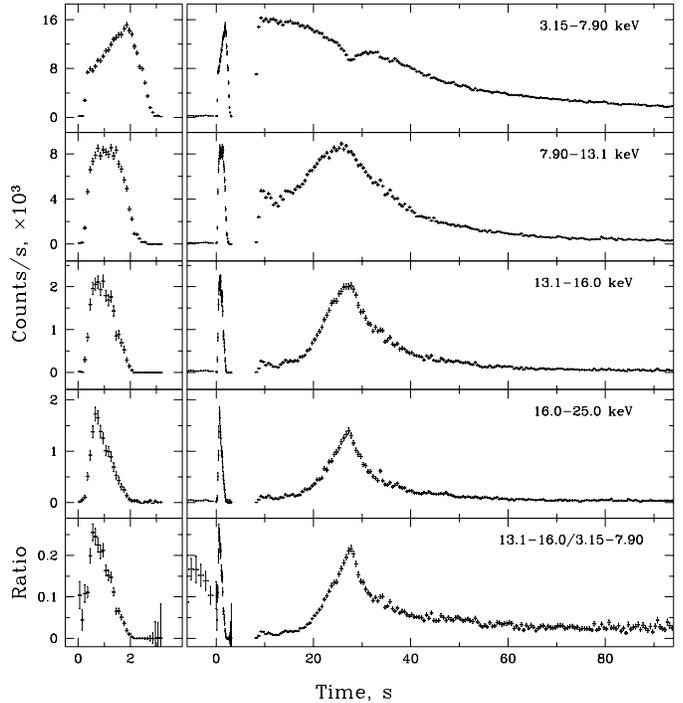}}

 \caption{\small Temporal profiles of the burst detected by RXTE in different
 energy bands and corresponding evolution of the hardness (ratio of
 fluxes in 13-16 and 3.1-7.9 keV bands). Left panels show profiles for
 the precursor with better time resolution.}  \label {lcurv}
\vspace{-1.4mm}
\end{figure}

Fig.\,\ref{totcur}$a$ shows the temporal profile of the burst in the
broad PCA energy band with 0.5-s time resolution. The moment ``0''
corresponds to 7:00:03 a.m. (UT). The burst began with an extremely
strong ($\sim 1.9$ Crab) precursor event which lasted $\sim 3$~s. Then a
2-s quiet interval was observed during which the flux fell by
$7.8\pm1.3$ times from the level measured before the burst. The drop is
well seen in the insertion to Fig.\,\ref{totcur}. A weak burst (the
second precursor) followed this time interval and gave a start for the
primary event. The second precursor was about 1.5-s long and a factor of
6 weaker than the first one. The primary event itself was $\sim 2$ Crab
in a peak and had rather a complex shape and a duration in excess of
150~s. In general the observed profile was very similar to that detected
by \mbox{ART-P} from \4u\ on February 17, 1992 (Fig.\,\ref{totcur}$b$).
The coincidence could be even better if the profiles obtained by PCA at
high $h\nu\ga 6$ keV energies (where ART-P was most sensitive) were used
(Fig.\,\ref{lcurv}). \mbox{ART-P} being a coded-mask telescope localized
its burst with a few arcmin accuracy and identified it with Terzan~2
(\cite{greb99}). The similarity in the two profiles allows us to
conclude that the burst detected by PCA occurred in the same source.

\begin{figure}[ht]
 \epsfxsize=89.5mm
 \hspace{-1.0mm}{\epsffile{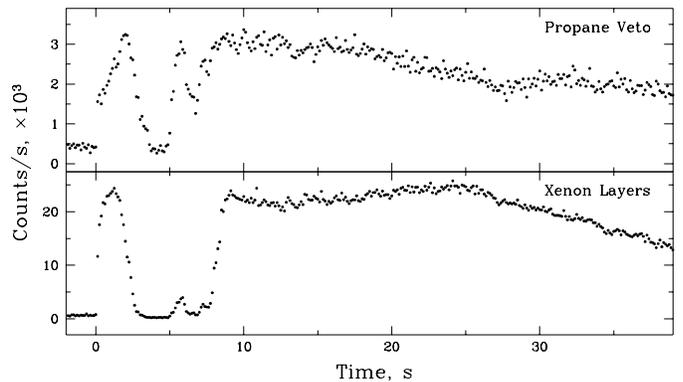}}

 \caption{\small Count rates taken from the PCA propane ($h\nu\leq 2$ keV) and
xenon ($h\nu\geq 2$ keV) anodes. During the second precursor the source
spectrum was dominated by very soft photons.}\label{propane}
\vspace{-3mm}
\end{figure}
In Fig.\,\ref{lcurv} we show the PCA count rate as a function of time in
four different energy bands. The moment ''0'' is the same as in
Fig.\,\ref{totcur}. The gap in the data in 3.25-7.75~s after the
precursor onset resulted from the Burst Catcher mode switching
off mentioned in sec.\,2.  The figure exhibits several notable features
in the burst and precursor profiles. It shows that although the first
rise in the flux occurred simultaneously and at nearly the same rate in
all bands, the time of reaching the maximum and the typical time
scale of the flux decline were different in the soft and hard bands.
During the precursor the peak flux was reached in $\sim1.9$ and $0.6$~s
after the onset in the \mbox{3.15-7.9} and 16-25 keV bands,
respectively.  Thus, the precursor was longer in the soft band. On the
contrary, during the primary event the rise in the soft flux was much
prompter and the peak was reached in $\sim9$ and 27~s in the same bands.
The peak observed at high energies was accompanied by a dip-like feature
in the 3.15-7.9 keV light curve.  Such behaviour reflected strong
changes in the source hardness occurred during the burst. The bottom
panel in Fig.\,\ref{lcurv} indicates that the maximum hardness was
observed twice -- during the precursor (in $\sim0.5$~s after the onset)
and during the primary event (in $\sim 27$~s). And it slowly decreased
during the decay of both events. Remarkably that the second precursor
was connected with the rise in the soft flux only. Fig.\,\ref{propane}
shows the burst profile taken from the propane anodes of PCA which are
sensitive to very soft ($h\nu\la2$~keV) photons. In this band the second
precursor was comparable in strength with the first one. The bottom
panel shows the hard X-ray profile from the xenon anodes (the same as in
Fig.\,\ref{totcur} but with better time resolution). Note that both the
profiles are given here without background subtraction.

A long exponential decay present in the profile beginning since
$\sim30$~s is a characteristic of the type I bursts. Using the data for
the first 60~s of the decay we estimated its time scale to be $\sim
21$~s at low energies and decreasing to $\sim8$~s at high energies. The
strong observed precursor comparable in a peak flux with the primary
event was on the contrary a completely unusual feature for \mbox{X-ray}
bursts. Although bursts with precursors were observed earlier
(\cite{lewin93}), \4u\ was likely the first source in which the
precursors were so powerful.
\vspace{-2mm}

\section{Spectral evolution during the burst}
For further analysis we got more than 100 consecutive photon
spectra with an integration time of 0.25~s (to cover the precursor and
the rise phase of the primary event) and 1~s (to cover the rest part
of the burst). The spectra were approximated in the \mbox{3-20} keV
band with the black body model. The interstellar absorption was taken
into account with $N_{\rm H}$ fixed at $1\times10^{22}\
\mbox{cm}^{-2}$ (as measured by ASCA, see \cite{barre99b}). The 
persistent spectrum was not subtracted. The drop in its flux by a factor
of 8 which followed the precursor has shown that the persistent source
was hidden by the expanding photosphere or suppressed by huge energy
release in the burst. It is unclear when the flux was able to recover
and what the level was. The HEXTE spectrometer which could clarify this
point had insufficient sensitivity at short time scales. We will see 
that neglect by the persistent emission was not crucial for our analysis
because of exceptional power of this burst.

\begin{figure}[t]
 \epsfxsize=89mm
 \hspace{-0mm}{\epsffile{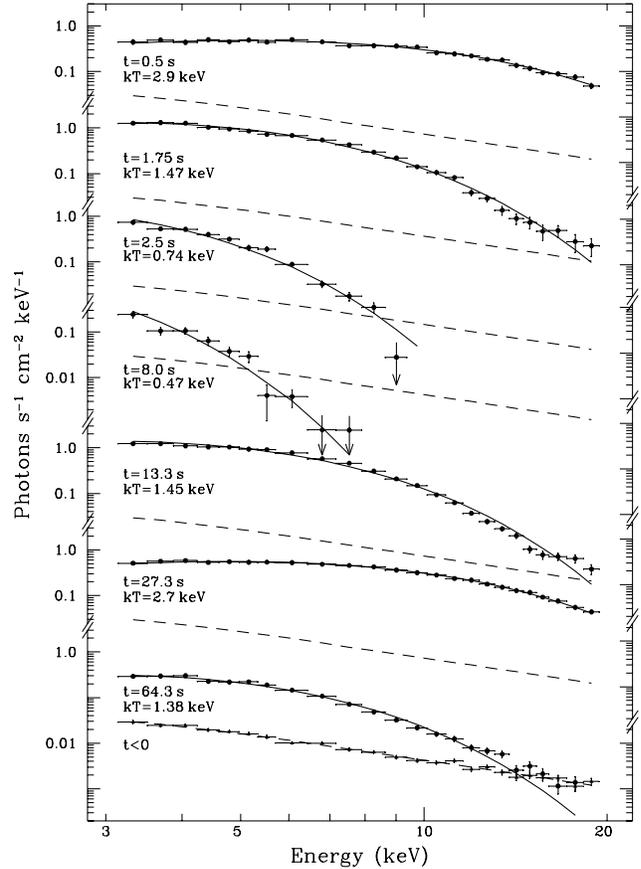}}

 \caption{\small Photon spectra measured at different phases of the burst
(filled cycles). The best-fit black body models are shown by solid
lines, the persistent spectrum taken before the burst -- by
triangles, its power law approximation -- by dashed lines. For all the
spectra the time since the burst onset and the black body temperature
are indicated. Error bars corresponds to $1\sigma$.} \label {spec}
\vspace{-2mm}
\end{figure}
\begin{figure}[t]
 \vspace{-0.4mm}
 \epsfxsize=96.5mm
 \hspace{-2mm}{\epsffile{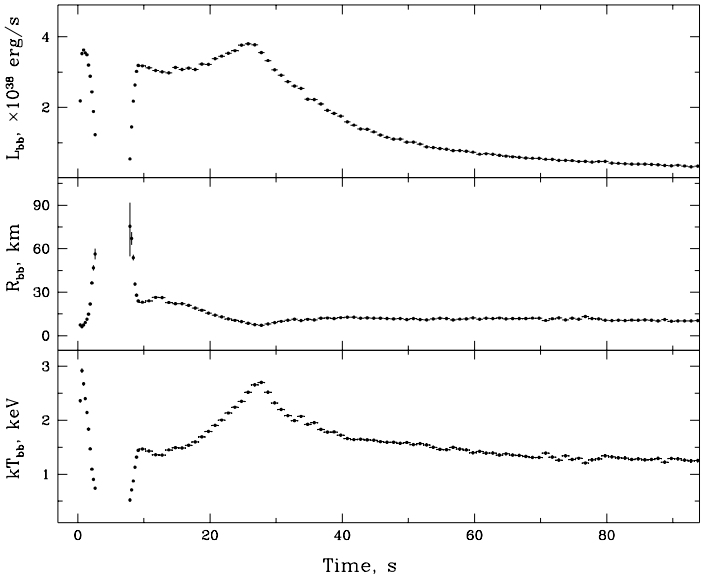}}

 \caption{\small Evolution of the bolometric luminosity, radius and effective
  temperature of the neutron star photosphere during the burst in
  \4u\ (from the black body fit to the spectra).} \label{param} 
\vspace{-3mm}
\end{figure}
Fig.~\ref{spec} presents several typical spectra of the source at
different phases of the burst. The best-fit approximation is shown by
solid lines.  The persistent spectrum accumulated during 16~s before the
burst and adequately fitted by the absorbed power law model with the
photon index $\alpha\simeq-1.97\pm0.03$ is shown by triangles in the
bottom panel and indicated by dashed lines in the other panels.  Note
that the spectrum measured at $t\simeq 8$~s exhibits the lack of X-rays
at $h\nu\ga5$ keV in comparison with the persistent one.  Note also that
the spectrum measured at $t\simeq64$~s, in the far tail of the burst,
points on the persistent emission recovery and its contribution to the
flux at $h\nu\ga15$ keV. The spectrum obtained in 300~s after the onset
was already very coincident with that observed before the burst.

The figure explains the evolution of the source hardness described in
sec.~3. Just after the onset (in 0.5~s) the source had a hard spectrum
corresponding to the black body temperature $kT\simeq2.9$ keV. The
temperature fell by 4 times during the next 2~s. After the gap in
the PCA data, $kT$ again was found to be small $\simeq0.47$ keV but it
quickly rose and reached $\simeq2.7$ keV in $27$~s after the onset.  The
gradual decrease of $kT$ followed this maximum and continued up to the
end of the observation.  The detailed information on the evolution of
spectral parameters ($kT$ and $R$ -- the black body radius of the
neutron star photosphere) is presented in Fig.\,\ref{param}.  In the
same figure we show the source bolometric luminosity
$L=4\pi R^{2}\, \sigma\, T^{4}$ ($\sigma$ is the Stefan-Boltzmann
constant) as a function of time. The distance to Terzan~2 was taken to
be 6.6~kpc (\cite{barbuy98}).

\vspace{-1.5mm}

\section{Discussion}

Fig.\,\ref{param} shows that the initial ($\sim 30$~s) stage of the
burst was connected with strong photospheric expansion (during the
precursor) and contraction (during the primary event). The observed
radii were in excess of 70 km, velocities --  of $100\
\mbox{km\,s}^{-1}$. The peak value of 
the luminosity, $L_{\rm p}\simeq 3.7\times10^{38}\ \mbox{erg s}^{-1}$,
was equal to the Eddington limit for a $M_*\simeq1.4\ M_{\sun}$
neutron star and helium-rich material. While expanding, the photosphere
got cool, its spectrum softened and the fraction of photons with
energies within the PCA band decreased. That was the reason why the
first narrow precursor appeared in the burst profile.

\begin{figure}[t]
 \vspace{-0.6mm}
 \epsfxsize=78mm
 \hspace{10mm}{\epsffile{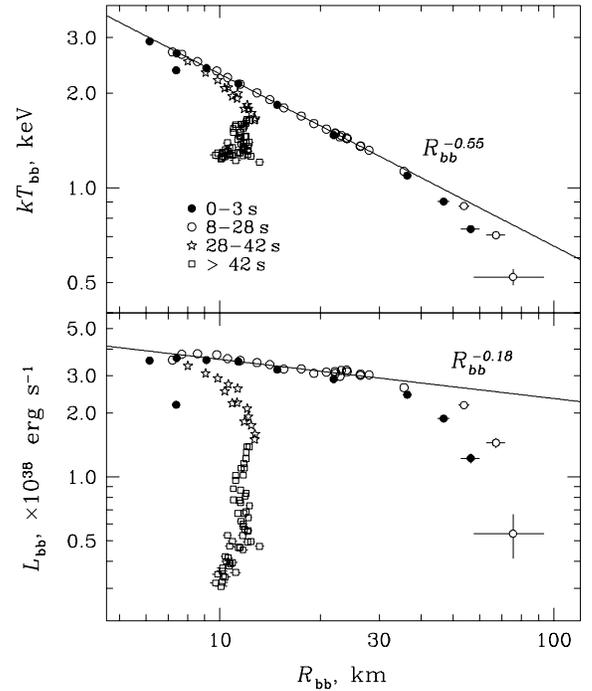}}

 \caption{\small Effective temperature and bolometric luminosity of the
  neutron star photosphere given as functions of its radius.}
 \label{depend} 
\end{figure}

The dependence of $T$ on $R$ during photospheric expansion and
contraction is shown in Fig.\,\ref{depend} by filled and open circles,
respectively. It could be described by a single law $T\sim R^{-\beta}$
with $\beta\simeq 0.546\pm0.003$ which implies that the luminosity
$L\sim T^4\,R^2\sim R^{-0.184}$ slowly decreased while the radius
increased in contrast with the usual suggestion that it remains constant
and very close to the Eddington limit (\cite{lewin93}). This issue may
however be hasty taking into account that our simple spectral modeling
ignored distortions produced in the spectrum by Comptonization in the
outer layers of the photosphere (\cite{ebisuzaki86,st86}).
Comptonization may be also responsible for the decrease in $R$ observed
in $0.5$ and 27 s after the onset (in comparison with the value
$R_*\simeq12$\,\,km found in the burst tail).

The total emitted energy $E\simeq 1.4\times 10^{40}\
\mbox{ergs},$  thus the mass of exploded matter $M\simeq
E/\epsilon_N\simeq 8\times10^{21}\ \mbox{g}$ where
$\epsilon_N\simeq0.002\,c^2$ is the efficiency of helium burning. The
Thomson depth of such matter $\tau_{\rm es}\simeq4\times10^6\,(R/100\,
\mbox{km})^{-2}\!.$ Really, only its small part could be involved in
expansion.  For a few seconds after the precursor, the photosphere,
extended but still opaque, completely hid the region in which the
gravitational energy of accreted matter was released.  The flux from the
source fell below the persistent level. It is unclear however whether
the accretion efficiency  was at the persistent level,
$\epsilon_G\simeq GM_*/R_*$, that time. The burst could destroy process of steady
accretion onto the source. The photospheric sound crossing time $R/v_s$
increased during expansion and could reach seconds if $R$ exceeded a few
$\times\ 10^2$ km.  Respectively, the second precursor in the burst
profile could be explained by photospheric oscillations.

Assuming $\epsilon_G$ to be $0.2\, c^2$ we can estimate the burst
recurrence time $t_{\rm r}\sim \epsilon_G E/\epsilon_N L_{\rm 
X}\simeq 2.3\times10^5\ \mbox{s}$. Here $L_X\simeq7\times10^{36}\,
\mbox{erg s}^{-1}$ is the persistent luminosity. The obtained  
value is 1.4 times smaller than the RXTE exposure.

\begin{acknowledgements} 
This research based on the data obtained through the HEASARC Online
Service. We thanks R.Sunyaev, N.Inogamov, M.Revnivtsev for useful
discussion and acknowledge support by RBRF grants 98-02-17056 and
99-02-18178.
\end{acknowledgements}


\end{document}